\documentclass{elsart}
\usepackage{amssymb}
\usepackage{amsmath}

\journal{\textbf{ International Journal of Engineering Science, 24},
 7, p.  1057-1066, \textbf{(1986).}\qquad \qquad\qquad\qquad\quad}
\input{tcilatex}
\begin{document}

\begin{frontmatter}

\title{Variational Principle Involving\\ the Stress Tensor in
Elastodynamics}
\author[France1]{Henri Gouin   }
\ead{henri.gouin@univ-cezanne.fr}

{\small and}

\author[France2]{\qquad \qquad\qquad\qquad\quad Jean-Fran\c cois
Debieve   }

\address[France1]{ C.N.R.S.  U.M.R.
6181  \&   University of Aix-Marseille\\
 Av. Escadrille Normandie-Niemen, Box 322, 13397 Marseille Cedex 20, France.  }
\address[France2]{  C.N.R.S.  U.M.R.  6595, IUSTI, \\  5 rue E. Fermi, 13453 Marseille Cedex 13 France.}


\address{}

\begin{abstract}
In the mechanics of inviscid conservative fluids, it is classical to
generate the equations of dynamics by formulating with adequate
variables, that the pressure integral calculated in the time-space
domain corresponding to the motion of the continuous medium is
stationary. The present study extends this principle to the dynamics
of large deformations for isentropic motions in thermo-elastic
bodies: we use a new way of writing the equations of motion in terms
of  \emph{potentials}  and we substitute the trace of the stress
tensor for the pressure term.

\end{abstract}


\end{frontmatter}

\section{Introduction}

{\footnotesize HAMILTON'S\ PRINCIPLE} holds for all conservative
mechanical systems with holonomic side conditions. This is the case
with perfect fluids in adiabatic motion.\ If $(F)$ represents the
set of the virtual motions which assign to the system two given
positions at times $t_{1}$ and $t_{2}$, the principle can be stated
as follows  (\emph{principle of least action}):

\textit{Among the }$(F)$\textit{-motions, the motion of the system is that
which stationarizes the integral of the difference between kinetic energy
and potential energy over the time-space domain occupied by the continuous
medium.}

Several authors have observed that in conventional fluid mechanics, many
variational principles can be reduced to Hamilton's principle, although the
procedures are diverse, according to the choice of the unknown functions. In
the Eulerian description of fluids, several authors [1-5] have shown that
the equations of motion may be derived from a variational principle through
the introduction of Lagrange multipliers corresponding to the side
conditions that the variations of the mass density, the entropy and the
Lagrangian coordinates are coupled with some conservation conditions. Minor
details apart, and with $\mathbf{V},\rho ,s,\mathbf{X},\phi ,\psi
,\varepsilon $ as independent variables, these methods consist in
stationarizing

\begin{eqnarray}
&&\int_{W}\left\{ \frac{1}{2}\rho \mathbf{V}^{2}-\rho \varepsilon -\rho
\Omega +\phi \left( \frac{\partial \rho }{\partial t}+\func{div}\rho \mathbf{%
V}\right) +\psi \left( \frac{\partial \rho s}{\partial t} +\func{div}\rho s%
\mathbf{V}\right) \right.  \notag \\
&&\qquad\left. + \left( \frac{\partial \rho \mathbf{X}^{\tau }}{\partial t}+%
\func{div}(\rho \mathbf{VX}^{\tau })\right) \mathbf{\Xi }~\right\} d
\emph{{\textbf{x}}}\,dt,
\end{eqnarray}%
where $\mathbf{X}$ $=%
\begin{vmatrix}
\,{X}^{1} \\
\,{X}^{2} \\
\,{X}^{3}%
\end{vmatrix}%
$ are the Lagrangian coordinates, $\textbf{\emph{x}}$ $=%
\begin{vmatrix}
\,{x}^{1} \\
\,{x}^{2} \\
\,{x}^{3}%
\end{vmatrix}%
$ the Eulerian coordinates, $\rho $ is the mass density, $\mathbf{V}$ the
velocity, $\varepsilon $ the internal energy density, $s$ the entropy
density, $\Omega $ the extraneous force potential; $\phi $ and $\psi $ are
scalar Lagrange multipliers, $\mathbf{\Xi }$ is a vector Lagrange multiplier
and $\,^\tau $ stands for the transposition. The variations with respect to $%
\phi, \psi, \mathbf{\Xi }\ $\ give the constraints:%
\begin{equation}
\frac{\partial \rho }{\partial t}+\func{div}\rho \mathbf{V}=0,\text{ \ \ \ \
\ \ density equation}
\end{equation}%
\begin{equation}
\frac{\partial \rho s}{\partial t}+\func{div}\rho s\mathbf{V}=0,\text{ \ \ \
\ \ entropy equation}
\end{equation}%
\begin{equation}
\frac{\partial (\rho \mathbf{X}^{\tau })}{\partial t}+\func{div}(\rho
\mathbf{VX}^{\tau })=0,\text{ \ \ \ \ \ Lin's constraint [5,6].}
\end{equation}%
The last equation expresses that, in the eulerian description, the initial
coordinates $\mathbf{X}$ do not change along a particle path [2]. The
connection of this principle with Hamilton's principle is somewhat intricate
for any direct use.\ Casal [7], Seliger \& Whitham [8] have introduced
considerable simplification. By integrating by parts, and supposing that the
integral corresponding to the boundary of the time-space domain has a
zero-variation, these authors reduce the integrand in (1) to:%
\begin{equation*}
\frac{1}{2}\rho \mathbf{V}^{2}-\rho \varepsilon -\rho \Omega -\rho \overset{%
\centerdot }{\phi }-\rho s\overset{}{\overset{\centerdot }{\psi }-\rho }%
\mathbf{X}^{\tau }\overset{\centerdot }{\mathbf{\Xi }}
\end{equation*}%
where $\ ^{\centerdot }\ $ is the material derivative. By using \emph{\
Clebsch representation} [9] for the motion and thermodynamic relations, they
reduce the Lagrangian density to be just the pressure $p$. Conversely they
have to find a suitable variational principle which generates the dynamical
equations of a fluid and auxiliary conditions by expressing that%
\begin{equation}
\int_{W}p~d\emph{\textbf{x}}\,dt
\end{equation}%
is stationary.

On this purpose, Seliger and Whitham express the pressure as a function of
the enthalpy and the entropy. They give a Clesch representation of the
velocity field by means of Lagrange multipliers $\phi ,\psi ,\mathbf{\Xi }.$
This does not restrict the form of the velocity vector and gives a redundant
number of unknowns.\ Under the constraint%
\begin{equation*}
p=\frac{1}{2}\rho \mathbf{V}^{2}-\rho \varepsilon -\rho \Omega -\rho\,
\overset{\centerdot }{\phi }-\rho s\,{\overset{\centerdot }{\psi }-\rho\, }%
\mathbf{X}^{\tau }\overset{\centerdot }{\mathbf{\Xi }},
\end{equation*}%
the variational principle which states that (5) is stationary with respect
to the variables $\phi ,s,\psi ,$ $\mathbf{X}$ and $\mathbf{\Xi }$ , yields
the equations of the fluid motion in terms of potentials, and the density
equation. \newline
Obviously, this principle is far from Hamilton's principle. Seliger and
Whitham noticed that Clesch's representation of the velocity is a crucial
step for the final form (5) and said \emph{this seems to be an especially
simple form limited to fluids}.

The purpose of the present study is to generalize this principle to
isentropic motions in thermo-elastic bodies with thermodynamic potentials
[10]. The medium may be inhomogeneous. \newline
We analyze the article by Seliger \& Whitham and the limitation caused by
Clesch representation. By using an appropriate representation for the
internal energy, we overcome the difficulties associated with the fact that
the stress tensor of an elastic medium is no more spherical.

In the second section, a decomposition of the stress tensor into spherical
and deviatoric parts is associated to partial derivations of the internal
energy. {\emph{The decomposition we obtained in 1986 is now largely used in
the literature}}.

In the third section, an Eulerian approach of the variational principle
similar to the method used by Seliger and Whitham leads to a new Clesch
representation and to the equations of motion in terms of potentials.\

In the fourth section, we observe that the Lagrangian density can be reduced
to some expression of the trace of the stress tensor.

We extend the results of Seliger and Whitham by writing that $\displaystyle%
\int_{W}tr\,\sigma ~d\emph{\textbf{x}}\, dt$ is stationary, where
$tr\,\sigma $ stands for the trace of the stress tensor. Now, we
consider $tr\,\sigma $ as
a function of the enthalpy $h$, the entropy $s$, the material variable $%
\mathbf{X}$ and of the tensor $C$ where $C$ is the right Cauchy-Green tensor.

The variations are taken with respect to variables $s,$ $\mathbf{X}$ $,\phi
,\psi ,\mathbf{\Xi }$ introduced in the same way as in relation (5) and
under the constraint:%
\begin{equation*}
h-\frac{1}{2}\mathbf{V}^{2}+\overset{\centerdot }{\phi }+\psi \overset{%
\centerdot }{s}-\mathbf{X}^{\tau }\overset{\centerdot }{\mathbf{\Xi }}%
+\Omega =0.
\end{equation*}%
Three appendices present auxiliary calculations and, by means of the
convective derivation associated with the velocity field, show the
equivalence between Euler equations and our new motion equations in terms of
potentials for isentropic processes.

\section{A decomposition of the stress tensor in a hyperelastic medium}

Each particle of the continuous medium is labelled by a material variable {$%
\mathbf{X}$}, ranging over a reference configuration $\mathcal{D}_{0}$ in an
Euclidian space [10]. The reference density $\rho _{0}$ is given as a
function on $\mathcal{D}_{0}$ [11]. \newline
The expression ${\mathbf{x}} =\phi (\mathbf{X},t)$ of the spatial position
describes the motion of the continuous medium.\ Generally, $\phi (\centerdot
,t)$ is a twice continuously differentiable diffeomorphism of $\mathcal{D}%
_{0}$ onto a compact orientable manifold $\mathcal{D}_{t}$ constituting the
image of the material at time $t$ (see Appendix 1). As usual, we denote by $F
$ the deformation gradient; then, $C=F^{\tau }F$ is the right Cauchy-Green
tensor. \newline
Recall that:%
\begin{equation}
\rho\ (\det C)^{\frac{1}{2}} =\rho _{0}(\mathbf{X}).
\end{equation}%
The internal energy density is supposed to be a function of the tensor $C$,
the specific entropy $s$ and the material variable $\mathbf{X}$ (the
hyperelastic medium is not necessary materially homogeneous):%
\begin{equation*}
\varepsilon =e(C,s,\mathbf{X}).
\end{equation*}%
In the case of an isentropic process, this leads us back to a medium
constituted with hyperelastic points (see Ref.10, p.19). Now, let us put:%
\begin{equation*}
C^{\prime }=\frac{1}{(\det C)^{\frac{1}{3}}}C.
\end{equation*}%
Then, $\displaystyle C= \left(\frac{\rho_0(\mathbf{X})}{\rho}\right)^{\frac{2%
}{3}} C^{\prime} $ and  the internal energy density can be expressed
in
the form:%
\begin{equation}
\varepsilon =f(\rho ,C^{\prime },s,\mathbf{X}).
\end{equation}%
When $f$ is independent of both $C^{\prime }$ and $\mathbf{X}$, we retrieve
the case of an elastic materially homogeneous fluid. \newline
Let us observe that the independent variables $\rho $ and $C^{\prime }$ are
substituted to $C$. Since essentially $\det C^{\prime }=1$, the variable $%
\rho $ corresponds to the change of volume while the tensorial variable $%
C^{\prime }$ represents the distorsion of the medium. This point is
fundamental for the decomposition of the stress tensor and will be the key
of the representation of the equations of motion in terms of  \emph{%
potentials}. But $f$ is defined on the manifold $\det C^{\prime }=1$ and it
is more convenient to introduce the function $g$ such that%
\begin{equation}
\varepsilon =g(\rho ,C,s,\mathbf{X})=f(\rho ,\frac{C}{(\det C)^{\frac{1}{3}}}%
,s,\mathbf{X})
\end{equation}%
where $g$ is a homogeneous function of degree zero with respect to $C$. We
assume $g$ to be a differentiable function of $\rho, C, s$, $\mathbf{X}$.
The stress tensor can be written [11,12]:%
\begin{equation}
\sigma =2\rho F\frac{\partial e}{\partial C}F^{\tau }.
\end{equation}%
From (6) and (8), it follows:%
\begin{equation*}
\sigma =2\rho F\left( \frac{\partial g}{\partial \rho }\frac{\partial \rho }{%
\partial C}+\frac{\partial g}{\partial C}\right) F^{\tau }.
\end{equation*}%
Using Eq. (27) proved in Appendix 2, we deduce:%
\begin{equation}
\sigma =-\rho
{{}^2}%
\frac{\partial g}{\partial \rho }I+2\rho F\frac{\partial g}{\partial C}%
F^{\tau },
\end{equation}%
where $I$ represents the identity tensor.\ Hence, with (10) and (29), we
deduce:%
\begin{equation*}
\sigma =-pI+\sigma _{1}
\end{equation*}%
with:%
\begin{equation}
\begin{array}{ccccccc}
\displaystyle p=\rho
{{}^2}%
\displaystyle\frac{\partial g}{\partial \rho }, &  & \displaystyle\sigma
_{1}=2\rho F\frac{\partial g}{\partial C}F^{\tau } &  & \text{and} &  & tr\,
\sigma _{1}=0.%
\end{array}%
\end{equation}

\bigskip

\section{A new transformation of the Hamilton principle for isentropic
processes in elastodynamics}

The Lagrangian density in the classical principle of Hamilton has the form
[2]:%
\begin{equation*}
L=\rho (\frac{1}{2}\mathbf{V}%
{{}^2}%
-\varepsilon -\Omega ).
\end{equation*}%
We introduce the variables $C$ and $s$ by means of (8) and the various
quantities $\rho ,\mathbf{V},C,s$ and $\Omega $ involved in the Lagrangian
density become functions of $\mathbf{X}$ and of the motion $\phi $ of the
continuous medium [2].

Let us denote by $\delta $, the virtual displacement as defined by Serrin
(see Ref. 2, p. 145). Under the Eulerian form, the variational principle of
Hamilton reads:

For all variations $\delta ,$ vanishing on the boundary of $W$, one has%
\begin{equation*}
\delta \int_{W}L~d\emph{\textbf{x}}\,dt=0.
\end{equation*}%
Consequently, equations of the motion of isentropic processes in
elastodynamics are obtained.

In another way, it is convenient to consider the quantities $\rho ,\mathbf{V}%
,s$ and $\mathbf{X}$ as variables. To be able to take into account the
constraints imposed on $\rho ,s$ and $\mathbf{X}$, it is necessary to
introduce appropriate Lagrange multipliers [2,7,8]\,($C$ being a function of
$\mathbf{X}$ by its derivative with respect to $\mathbf{X}$). We are led to
the variational principle:

\begin{theorem}
Let $\phi ,\psi $ and $\mathbf{\Xi }$ be three Lagrange multipliers where $%
\phi ,\psi $ are scalars and $\mathbf{\Xi }$ is a vector, the conditions for%
\begin{equation}
\int_{W}\left\{ \rho (\frac{1}{2}\mathbf{V}%
{{}^2}%
-\varepsilon -\Omega )+\phi (\frac{\partial \rho }{\partial t}+\func{div}%
\rho \mathbf{V})-\rho \psi \overset{\centerdot }{s}-\rho \mathbf{\Xi }^{\tau }%
\overset{\centerdot }{\mathbf{X}}\right\} ~d{\it{\textbf x}}\,dt=0
\end{equation}%
to be stationary for every variation of $\rho ,\mathbf{V},s$,$\textbf{X}$$%
,\phi ,\psi ,\mathbf{\Xi }$ vanishing on the boundary of $W$ yield
the equations of isentropic motion and the constraints (2)-(4).
\end{theorem}

The Lagrange multiplier $\mathbf{\Xi }$ is associated with Lin's constraint:%
\begin{equation*}
\frac{\partial }{\partial t}(\rho \mathbf{X}^{\tau })+\func{div}(\rho
\mathbf{VX}^{\tau })=0.
\end{equation*}%
The latter arises from the fact that the Lagrangian coordinates no longer
need to be given by explicit expressions; only the velocity field $\mathbf{V}
$ must be such that it should be possible to obtain such a coordinate system
by integration [5,7,8].

By integrating by parts the two expressions in integral (12)
\begin{equation*}
\begin{array}{ccccc}
\displaystyle\phi \left( \frac{\partial \rho }{\partial t}+\func{div}\rho
\mathbf{V}\right) &  & \text{and} &  & -\rho\, \mathbf{\Xi }^{\tau }\overset{%
\centerdot }{\mathbf{X}},%
\end{array}%
\end{equation*}%
with zero-variation for the terms on the boundary of$\ W$, we are reduced to
an equivalent theorem:

\begin{theorem}
With $\rho, \mathbf{V}, s$, $\mathbf{X}$$,\phi, \psi, \mathbf{\Xi }$
as
independent unknown functions with variations vanishing on the boundary of $%
W $, and%
\begin{equation}
\Lambda =\rho (\frac{1}{2}\mathbf{V}%
{{}^2}%
-\varepsilon -\Omega -\overset{\centerdot }{\phi }-\psi\, \overset{\centerdot }{s}+%
\mathbf{X}^{\tau }\overset{\centerdot }{\mathbf{\Xi }})
\end{equation}%
the conditions for%
\begin{equation}
\int_{W}\Lambda ~d{\it{\textbf x}}\,dt
\end{equation}%
to be stationary yield the equations of isentropic motion in
elastodynamics and the constraints (2)-(4).
\end{theorem}

Proof of the theorem:

In the last part of this paper, $\displaystyle\frac{\partial
}{\partial \emph{\textbf{x}}}$ will be used to denote the gradient
 associated with a tensor quantity defined on the motion
space $\mathcal{D}_{t}$ and ( \thinspace\ )' will be used to denote
the partial derivatives.\ Let us write $\theta =g_{s}^{\prime },$
the Kelvin temperature, and
\begin{equation*}
h=\varepsilon +\rho g_{\rho }^{\prime }.
\end{equation*}%
We call $h$, the scalar specific enthalpy density and shortly enthalpy in
this paper. The variations corresponding to $\mathbf{V},\rho ,s$ and $%
\mathbf{X}$ in expression (14) lead to the following equations: (see
Appendix 3)%
\begin{equation}
\begin{array}{ccc}
\delta \mathbf{V}: &  & \mathbf{V}^{\tau }\displaystyle=\frac{\partial \phi
}{\partial \emph{\textbf{x}}}+\psi \frac{\partial s}{\partial \emph{\textbf{x}}}-\mathbf{X}%
^{\tau }\frac{\partial \mathbf{\Xi }}{\partial \emph{\textbf{x}}}%
\end{array}%
\end{equation}%
\begin{equation}
\begin{array}{ccccccc}
\delta \rho : &  & \Lambda _{\rho }^{\prime }=0, &  & \text{hence:} &  & %
\displaystyle\frac{1}{2}\mathbf{V}%
{{}^2}%
-h-\Omega -\overset{\centerdot }{\phi }-\psi \overset{\centerdot }{s}+%
\mathbf{X}^{\tau }\overset{\centerdot }{\mathbf{\Xi }}=0%
\end{array}
\tag{15$^{2}$}
\end{equation}%
\begin{equation}
\begin{array}{ccc}
\delta s: &  & -\rho \,\theta +\rho \,\overset{\centerdot }{\psi }=0%
\end{array}
\tag{15$^{3}$}
\end{equation}%
\begin{equation}
\begin{array}{ccc}
\delta \mathbf{X}: &  & \overset{\centerdot }{\mathbf{\Xi }}=g_{\mathbf{X}%
}^{\prime \tau }+\displaystyle\frac{1}{\rho }\,[\func{div}(\sigma
_{1}F)]^{\tau }.%
\end{array}
\tag{15$^{4}$}
\end{equation}%
Then, the variations of $\phi ,\psi $ and $\mathbf{\Xi }$ give the
constraints:%
\begin{equation*}
\begin{array}{lllllll}
\displaystyle\frac{\partial \rho }{\partial t}+\func{div}\rho \mathbf{V}=0,
&  & \overset{\centerdot }{s}=0 &  & \text{and} &  & \overset{\centerdot }{%
\mathbf{X}}=0.%
\end{array}%
\end{equation*}%
By denoting $\lambda =\phi -\mathbf{\Xi }^{\tau }$$\mathbf{X}$$,$ we deduce
the equations in terms of potentials for the motion:%
\begin{equation*}
\mathbf{V}^{\tau }=\frac{\partial \lambda }{\partial
\emph{\textbf{x}}}+\psi \frac{\partial s}{\partial
\emph{\textbf{x}}}+\mathbf{\Xi }^{\tau }F^{-1}
\end{equation*}%
\begin{equation}
\begin{array}{lll}
\overset{\centerdot }{\lambda }=\displaystyle\frac{1}{2}\mathbf{V}%
{{}^2}%
-h-\Omega , &  & \qquad \displaystyle\frac{\partial \rho }{\partial t}+\func{%
div}\rho \mathbf{V}=0,\notag%
\end{array}%
\end{equation}%
\begin{equation}
\begin{array}{lll}
\overset{\centerdot }{\psi }=\theta , & \qquad  & \overset{\centerdot }{s}=0,%
\end{array}%
\end{equation}%
$
\qquad\qquad\quad\overset{\centerdot }{\mathbf{\Xi }^{\tau }} = \displaystyle g_{\mathbf{X}%
}^{\prime }+\frac{1}{\rho }\ [\func{div}(\sigma _{1}F)],\quad\quad\
\, \overset{\centerdot }{\mathbf{X}}=0. $

 In Appendix 1, we prove
directly the equivalence between system (16) and the Euler equations
joined with constraint requirements.

\section{ {A simplified form of variational principle}}

In the same way as Selinger and whitham, in the case of perfect
compressible fluid [8], we shall show that the form of the
variational principle given in Section 3 may be considerably
simplified. We obtain a similar result in the case of isentropic
motions for elastodynamics with large deformations.

Equation (13) may be written $\Lambda =\rho \Lambda _{\rho }^{\prime }+\rho
{{}^2}%
g_{\rho }^{\prime }$ hence $\Lambda =\rho \Lambda _{\rho }^{\prime }+p,$
where $p=-\displaystyle \frac{1}{3}\,tr\,\sigma .$ From (15$^{2}$), we
obtain for the actual motions of the medium:%
\begin{equation*}
\Lambda =p.
\end{equation*}%
The Legendre transformation of $\rho\, g$ with respect to $\rho $ gives $-p$
with $p$ being   a function of the variables $h, s, C, \mathbf{X},$%
\begin{equation*}
p=l(h,s,C,\mathbf{X}),
\end{equation*}%
where $l$ is a homogeneous function of degree zero with respect to $C.$

\textit{Let us examine a converse}. First, we notice that, for the
thermo-elastic points of the medium, the knowledge of $p$ as a function of
the enthalpy $h$, the entropy $s$, $C$ and $\mathbf{X}$, homogeneous of
degree zero with respect to $C$, permits the deduction by means of
conventional thermodynamic relations, of the values of mass density,
temperature and internal energy density. Then,%
\begin{equation}
\begin{array}{ll}
\rho =p_{h}^{\prime }, &
\end{array}%
\qquad \theta =-\frac{p_{s}^{\prime }}{p_{h}^{\prime }}
\end{equation}%
and%
\begin{equation}
\varepsilon =h-\frac{p}{p_{h}^{\prime }}.
\end{equation}%
By differentiating (18), we obtain,%
\begin{equation*}
d\varepsilon =dh-\frac{dp}{\rho }+\frac{p}{\rho
{{}^2}%
}\;d\rho .
\end{equation*}%
Or
\begin{equation*}
d\varepsilon =dh-\frac{1}{\rho }\;p_{h}^{\prime }\;dh-\frac{1}{\rho }\;
p_{s}^{\prime }\;ds-\frac{1}{\rho }\;p_{\mathbf{X}}^{\prime }\;d\mathbf{X}-%
\frac{1}{\rho }\;tr(p_{C}^{\prime }\;dC)+\frac{p}{\rho
{{}^2}%
}\;d\rho .
\end{equation*}%
From (17), we deduce:%
\begin{equation*}
d\varepsilon =\theta \;ds+\frac{p}{\rho
{{}^2}%
}\;d\rho -\frac{1}{\rho }\;p_{\mathbf{X}}^{\prime }d\mathbf{X}-\frac{1}{\rho
}\; tr(p_{C}^{\prime }dC).
\end{equation*}%
Recall that the internal energy density $\varepsilon $ is in the form:%
\begin{equation*}
\varepsilon =g(\rho ,C,s,\mathbf{X}),
\end{equation*}%
where $g$ is a homogeneous function of degree zero with respect to $C$. If
we choose $d$$\mathbf{X}$$=0,$ $ds=0$ and $dC=Cd\tau ,$ from relations $%
tr(p_{C}^{\prime }C)=0,$ $tr(g_{C}^{\prime }C)=0$, and (28) we obtain:%
\begin{equation*}
\rho
{{}^2}%
g_{\rho }^{\prime }=p
\end{equation*}%
(see Appendix 2).

We can write the thermodynamic relations%
\begin{equation*}
\begin{array}{lllllll}
\rho
{{}^2}%
g_{\rho }^{\prime }=p, &  & \rho g_{C}^{\prime }=-p_{C}^{\prime }, &  &
\theta =g_{s}^{\prime }, &  & \rho g_{\mathbf{X}}^{\prime }=-p_{\mathbf{X}%
}^{\prime }.%
\end{array}%
\end{equation*}%
The velocity vector may be written under a Clebsch representation:%
\begin{equation}
\mathbf{V}^{\tau }=\frac{\partial \phi }{\partial
\emph{\textbf{x}}}+\psi
\frac{\partial s}{\partial \emph{\textbf{x}}}-\mathbf{X}^{\tau }\frac{%
\partial \mathbf{\Xi }}{\partial \emph{\textbf{x}}},
\end{equation}%
where $\phi ,\psi $ are scalar and $\mathbf{\Xi }$ is a vector, constituting
evidently a redundant set of unknown functions.

As stated in the last principle, where the equation
$\displaystyle\frac{\partial
\rho }{\partial t}+\func{div}\rho \mathbf{V}=0$ specifies $\rho ,$ $\overset{%
\centerdot }{s}=0$ specifies $s$ and $\overset{\centerdot
}{\mathbf{X}}=0$ specifies \textbf{$X$}, we shall require $h$ to
satisfy
\begin{equation}
h=\frac{1}{2}\mathbf{V}%
{{}^2}%
-\overset{\centerdot }{\phi }-\psi \;\overset{\centerdot }{s}+\mathbf{X}%
^{\tau }\overset{}{\overset{\centerdot }{\mathbf{\Xi }}-\Omega }.
\end{equation}%
(This constraint is connected with partial result (15$^{2}$)). We may write:

\bigskip

\begin{theorem}
For every variation of $s,$$\textbf{X}$$,\psi ,\phi ,\mathbf{\Xi }$
submitted to%
\begin{equation*}
h-\frac{1}{2}\mathbf{V}%
{{}^2}%
+\overset{\centerdot }{\phi }+\psi\; \overset{\centerdot }{s}-\mathbf{X}^{\tau }\overset{%
}{\overset{\centerdot }{\mathbf{\Xi }}+\Omega }=0
\end{equation*}%
and vanishing on the boundary of $W$, the conditions for
$\displaystyle\int_{W}tr\,\sigma ~d{\it{\textbf x}}\,dt$ to be
stationary yield the equations of isentropic motions in
elastodynamics and the relations (2)-(4)  (the variations of $C$ are
deduced from the variations of $\mathbf{X}$ by means of (30)). In
the same way as in
paragraph 3, we obtain  System (21):%
\begin{equation}
\begin{array}[t]{lll}
\delta \phi : &  & \displaystyle\frac{\partial \rho }{\partial t}+\func{div}\rho \mathbf{V%
}=0,  \\
\delta \psi : &  & \overset{\centerdot }{s}=0 ,\\
\delta s \,: &  & \overset{\centerdot }{\psi }=\theta, \\
\delta \mathbf{\Xi} : &  & \overset{\centerdot }{\mathbf{X}}=0,  \\
\delta \mathbf{X} : &  & \overset{\centerdot }{\mathbf{\Xi }^{\tau }}%
=g_{\mathbf{X}}^{\prime }+\displaystyle\frac{1}{\rho }\;\func{div}(\sigma _{1}F).%
\end{array}%
\end{equation}%
The set Eqs. (19-20) and System (21) constitutes a system of
equations of the motion and side conditions.
\end{theorem}

In this new form of variational principle, constraints are different from
those of Hamilton's principle. The result is that Lagrangian and independent
parameters are completely changed. \newline
Let us notice that with the new parameter $\lambda =\phi -\mathbf{\Xi }%
^{\tau }$$\mathbf{X}$ and with a Clebsch representation for the velocity
vector in the form%
\begin{equation*}
\mathbf{V}^{\tau }=\frac{\partial \lambda }{\partial
\emph{\textbf{x}}}+\psi \frac{\partial s}{\partial
\emph{\textbf{x}}}+\mathbf{\Xi }^{\tau }F^{-1},
\end{equation*}%
we may write:

\bigskip

\begin{theorem}
For every variation of $s,$$\textbf{X}$$,\psi ,\lambda ,\mathbf{\Xi
}$
submitted to%
\begin{equation*}
h-\frac{1}{2}\mathbf{V}%
{{}^2}%
+\overset{\centerdot }{\lambda }+\psi\, \overset{\centerdot }{s}+\Xi ^{\tau }\overset{%
\centerdot }{\mathbf{X}}+\Omega =0
\end{equation*}%
and vanishing on the boundary of $W$, the conditions for
$\displaystyle\int_{W}tr\,\sigma ~d{\it{\textbf x}}\,dt$ to be
stationary yield the equations of isentropic motions and the
relations (2)-(4).
\end{theorem}

\section{Conclusion}

The variational statement in Theorem 4 is far from Hamilton's principle .
The resulting motion equations involve thermodynamic variables like
temperature and entropy.\newline
The constraint condition $h-\displaystyle\frac{1}{2}\mathbf{V}%
{{}^2}%
+\overset{\centerdot }{\lambda }+\psi\,\overset{\centerdot }{s}+\Xi ^{\tau }%
\overset{\centerdot}{\mathbf{X}}+\Omega =0$ involves the specific enthalpy.
Although this constraint explicitly uses entropy and the non uniquely
defined parameters $\lambda $ and $\psi ,$ the variational principle in this
form is interesting in view of the fact that the stress tensor $\sigma $ has
an experimental meaning.

\text{\textbf{Acknowledgements}}

{\small We are grateful to Professor P. Casal for his incisive and
illuminating criticism. Partial support of this research (\emph{H.G.}) was
provided by DGA/DRET-France under contract 82-455.}

\vskip 1cm

\section{Appendix 1}

\textbf{Equations of isentropic motions in elastodynamics}

\textit{\qquad Preliminaries} [13,14]

The motion of the medium consists in the $t$-dependent $C^{2}$-diffeomorphism%
\begin{equation*}
\begin{array}{lllll}
\mathbf{X}=%
\begin{vmatrix}
\,X^{1} \\
\,X^{2} \\
\,X^{3}%
\end{vmatrix}%
\in \mathcal{D}_{0} & \overset{\phi _{t}}{\longrightarrow } & \emph{\textbf{x}}=%
\begin{vmatrix}
\,x^{1} \\
\,x^{2} \\
\,x^{3}%
\end{vmatrix}%
\in \mathcal{D}_{t}, &  & [11,12].%
\end{array}%
\end{equation*}%
We immediately deduce $\overset{\centerdot }{F}=\displaystyle\frac{\partial
\mathbf{V}}{\partial \emph{\textbf{x}}}F$ and $\overset{\centerdot }{%
\widehat{F^{-1}}}=-F^{-1}\displaystyle\frac{\partial
\mathbf{V}}{\partial \emph{\textbf{x}}}$. Let us write $T^{\ast
}(\mathcal{D}_{t})$ for the cotangent fiber bundle of
$\mathcal{D}_{t}$ and $T_{\emph{\textbf{x}}}^{\ast
}(\mathcal{D}_{t})$ the cotangent linear space to $\mathcal{D}_{t}$ at $%
\emph{\textbf{x}}$; then,%
\begin{equation*}
\begin{array}{ccc}
\emph{\textbf{x}}\in \mathcal{D}_{t} & \longrightarrow & L(t,\emph{\textbf{x}}%
)\in T_{\emph{\textbf{x}}}^{\ast }(\mathcal{D}_{t})%
\end{array}%
\end{equation*}%
represents a differential form field on $\mathcal{D}_{t}.$

Let us write $T^{\ast }(\mathcal{D}_{0})$ the cotangent fiber bundle of $%
\mathcal{D}_{0}$ and $T_{\mathbf{X}}^{\ast }(\mathcal{D}_{0})$ the cotangent
linear space to $\mathcal{D}_{0}$ at $\mathbf{X}$. The mapping $\phi
_{t}^{\ast }$ is induced by $\phi _{t}$ for the form fields. The convective
derivation $d_{c}$ of a form field $L$ is deduced from the diagram:%
\begin{equation}
\begin{array}{ccc}
L\in T^{\ast }(\mathcal{D}_{t}) & \overset{\phi _{t}^{-1^{\ast }}}{%
\longrightarrow } & LF\in T^{\ast }(\mathcal{D}_{0}) \\
\downarrow d_{c} &  & \downarrow \frac{d}{dt} \\
\overset{\centerdot }{L}+L\displaystyle\frac{\partial
\mathbf{V}}{\partial \emph{\textbf{x}}}\in T^{\ast
}(\mathcal{D}_{t}) & \overset{\phi _{t}^{^{\ast }}}{\longleftarrow }
& \overset{\centerdot }{LF}+L\displaystyle
\frac{\partial \mathbf{V}}{\partial \emph{\textbf{x}}}F\in T^{\ast }(%
\mathcal{D}_{0})%
\end{array}%
\end{equation}%
where $\overset{\centerdot }{L}+L\displaystyle\frac{\partial \mathbf{V}}{%
\partial \emph{\textbf{x}}}$ is the Lie derivative of $L$ with respect to
the velocity field $\mathbf{V}$ that is the infinitesimal
transformation of the one-parameter group of transformations $\phi
_{t}$ [15].

\textit{\qquad Consequences}

Let $b$ be a scalar field on $\mathcal{D}_{t}$ assumed to be an
Euclidian space; $\func{grad}$ represents the gradient operator on
$\mathcal{D}_{t}.$
We define two form fields by their values $\mathbf{V}^{\tau }$ and $(\func{%
grad}b)^{\tau }$. From (22) we deduce:%
\begin{equation}
d_{c}(\mathbf{V}^{\tau })=\mathbf{\Gamma }^{\tau }+\frac{\partial }{\partial
\emph{\textbf{x}}}(\frac{1}{2}\mathbf{V}%
{{}^2}%
),
\end{equation}%
(where $\mathbf{\Gamma }$ is the acceleration), and%
\begin{equation}
d_{c}(\func{grad}b)^{\tau }=(\func{grad}\overset{\centerdot }{b})^{\tau }
\end{equation}%
\begin{equation}
d_{c}(LF^{-1})=\overset{\centerdot }{L}F^{-1}.
\end{equation}

\textit{\qquad Potential equations}

With the notations of Section 2, the motion equation is:%
\begin{equation}
\rho \mathbf{\Gamma }^{\tau }+\frac{\partial p}{\partial \emph{\textbf{x}}}-%
\func{div}\sigma _{1}+\rho \frac{\partial \Omega }{\partial\emph{\textbf{x}}}%
=0.
\end{equation}%
Relation (8) gives:%
\begin{equation*}
d\varepsilon =\theta\, ds+\frac{p}{\rho
{{}^2}%
}\,d\rho +g_{\mathbf{X}}^{\prime }\,d\mathbf{X}+tr(g_{C}^{\prime }dC).
\end{equation*}%
By symmetry property of the tensor $g_{C}^{\prime },$ we obtain:%
\begin{equation*}
tr(g_{C}^{\prime }dC)=2tr(g_{C}^{\prime }F^{\tau }dF),
\end{equation*}%
and Eq. (11) implies%
\begin{equation*}
tr(g_{C}^{\prime }dC)=\frac{1}{\rho }\,tr(\sigma _{1}dFF^{-1}).
\end{equation*}%
From the definition of the specific enthalpy, we deduce:%
\begin{equation*}
\func{grad}h=\func{grad}\varepsilon +\frac{1}{\rho }\func{grad}p-\frac{p}{%
\rho
{{}^2}%
}\func{grad}\rho,
\end{equation*}%
\begin{equation*}
\frac{\partial h}{\partial \emph{\textbf{x}}}=\theta \frac{\partial s}{%
\partial \emph{\textbf{x}}}+\frac{1}{\rho }\frac{\partial p}{\partial
\emph{\textbf{x}}}+g_{\mathbf{X}}^{\prime }F^{-1}+\frac{1}{\rho }\func{div}%
(\sigma _{1}F)F^{-1}-\frac{1}{\rho }\func{div}\sigma _{1}.
\end{equation*}%
Equation (26) becomes:%
\begin{equation*}
\rho \mathbf{\Gamma }^{\tau }+\rho \frac{\partial }{\partial \mathbf{\emph{x}%
}}(h+\Omega )-\func{div}(\sigma _{1}F)F^{-1}-\rho\, \theta\, \frac{\partial s%
}{\partial \mathbf{\emph{x}}}-g_{\mathbf{X}}^{\prime }F^{-1}=0.
\end{equation*}%
Equation (23) implies:%
\begin{equation*}
d_{c}(\mathbf{V}^{\tau })=\frac{\partial }{\partial \emph{\textbf{x}}}(\frac{%
1}{2}\mathbf{V}%
{{}^2}%
-h-\Omega )+\theta \frac{\partial s}{\partial \emph{\textbf{x}}}+\frac{1}{%
\rho }\,\{ \func{div}(\sigma _{1}F)+\rho g_{\mathbf{X}}^{\prime
}\}F^{-1}.
\end{equation*}%
Let $\mathbf{\Xi }^{\tau }$ be a form field such that $\overset{\centerdot }{%
\mathbf{\Xi }^{\tau }}=\displaystyle\frac{1}{\rho }\,\func{div}(\sigma
_{1}F)+g_{\mathbf{X}}^{\prime }$. With substituting from Eq. (25),%
\begin{equation*}
d_{c}(\mathbf{V}^{\tau }-\mathbf{\Xi }^{\tau }F^{-1})=\frac{\partial }{%
\partial \emph{\textbf{x}}}\left(\frac{1}{2}\mathbf{V}%
{{}^2}%
-h-\Omega \right)+\theta \frac{\partial s}{\partial
\emph{\textbf{x}}}.
\end{equation*}%
Let $\lambda $ and $\psi $ be two scalar fields verifying $\overset{%
\centerdot }{\lambda }=\displaystyle\frac{1}{2}\mathbf{V}%
{{}^2}%
-h-\Omega $ and $\overset{\centerdot }{\psi }=\theta $. By adding to $%
\mathbf{\Xi }^{\tau }$, if necessary, an appropriate one-form with a zero
convective derivation, Eq. (24) and $\overset{\centerdot }{s}=0$ (isentropic
process), give:%
\begin{equation*}
\mathbf{V}^{\tau }=\frac{\partial \lambda }{\partial
\emph{\textbf{x}}}+\psi \frac{\partial s}{\partial
\emph{\textbf{x}}}+\mathbf{\Xi }^{\tau }F^{-1},
\end{equation*}%
where%
\begin{equation*}
\begin{array}{lllllll}
\overset{\centerdot }{\lambda }=\displaystyle\frac{1}{2}\mathbf{V}%
{{}^2}%
-h-\Omega , &  & \overset{\centerdot }{\psi }=\theta &  & \text{and}
&  & \overset{\centerdot }{\mathbf{\Xi }}^{\tau
}=\displaystyle\frac{1}{\rho }\,
\func{div}(\sigma _{1}F)+g_{\mathbf{X}}^{\prime }.%
\end{array}%
\end{equation*}%
These equations along with the relations (2), (3), (4) are potential
equations of the motion such as have been obtained in paragraph 3.

\section{Appendix 2}

\textbf{Some calculus}

\qquad \textit{Mass density}

By differentiating Eq. (6) and using Jacobi's identity, we obtain:%
\begin{equation*}
d\rho =-\frac{1}{2}\rho\, tr(C^{-1}dC)+(\det C)^{-\frac{1}{2}}\,\frac{%
\partial \rho }{\partial \mathbf{X}}\,d\mathbf{X}.
\end{equation*}%
Hence,%
\begin{equation}
\rho _{C}^{\prime }=-\frac{1}{2}\rho\, C^{-1}.
\end{equation}%
If we choose $dC=Cd\tau $ and $d$$\mathbf{X}$$=0$, this reduces to%
\begin{equation}
d\rho =-\frac{3}{2}\,\rho\, d\tau .
\end{equation}%
With respect with $C,$ $g$ is a homogeneous function of degree zero.\ We
deduce immediately (Euler identity):
\begin{equation}
tr(g_{C}^{\prime }C)=0.
\end{equation}

\qquad \textit{Cauchy-Green tensor}

Let us calculate the Cauchy-Green tensor variations:%
\begin{equation}
\begin{array}{lll}
dC=dF^{\tau }F+F^{\tau }dF, &  & \delta F=\displaystyle-F\frac{\partial
\delta \mathbf{X}}{\partial \emph{\textbf{x}}}F.%
\end{array}%
\end{equation}%
Hence,%
\begin{equation}
\delta C=-\left[C\frac{\partial \delta \mathbf{X}}{\partial \emph{\textbf{x}}}F+(C%
\frac{\partial \delta \mathbf{X}}{\partial
\emph{\textbf{x}}}F)^{\tau }\right].
\end{equation}

\section{Appendix 3}

\textbf{\ Proof of Eqs.(15)}

Let us write (14) under the form:%
\begin{eqnarray*}
\int_{W}\Lambda ~d\mathbf{\emph{x}} dt &=&\int_{W} \rho\left\{ \frac{1}{2}%
\mathbf{V}%
{{}^2}%
-\varepsilon -\Omega -\frac{\partial \phi }{\partial t}-\frac{\partial \phi
}{\partial \emph{\textbf{x}}}\mathbf{V}-\psi \frac{\partial s}{\partial t}%
\right. \\
&&\qquad\qquad\left. -\psi \frac{\partial s}{\partial \emph{\textbf{x}}}%
\mathbf{V}+\mathbf{X}^{\tau }\frac{\partial \mathbf{\Xi }}{\partial t}+%
\mathbf{X}^{\tau }\frac{\partial \mathbf{\Xi }}{\partial \emph{\textbf{x}}}%
\mathbf{V}\right\} ~d\emph{\textbf{x}}\,dt.
\end{eqnarray*}
For every $\delta \mathbf{V}$-variation, we deduce:%
\begin{equation*}
\int_{W}\rho \left\{ \mathbf{V-}\frac{\partial \phi }{\partial \emph{\textbf{x}}}-
\psi \frac{\partial s}{\partial \emph{\textbf{x}}}+\mathbf{X}^{\tau }%
\frac{\partial \mathbf{\Xi }}{\partial \emph{\textbf{x}}}\right\}
~\delta \mathbf{V}d\emph{\textbf{x}}\,dt=0,
\end{equation*}%
and we obtain immediately Eq. (15). An integration by parts, on the
boundary of $W$, leads with the same calculation to Eqs.
(15$^{2}$),(15$^{3}$) and the constraint relations.

For every $\delta \mathbf{X}$-variation, vanishing on the boundary of $W$,
we obtain:%
\begin{equation*}
\int_{W}\left\{ -\rho\, g_{\mathbf{X}}^{\prime }\delta
\mathbf{X}+\rho\, \overset{\centerdot }{\mathbf{\Xi }}^{\tau }\,
\delta \mathbf{X}-\rho\, tr(g_{C}^{\prime }\delta C)\right\}
~d\emph{\textbf{x}}\,dt=0,
\end{equation*}%
where $\delta C$ is a function of $\delta $$\mathbf{X}$ (see
Appendix 2, Eq.(30)).

By using Eqs.(11), (30), we obtain:%
\begin{equation*}
-\rho\, tr(g_{C}^{\prime }\delta C)=tr\left( \sigma
_{1}F\frac{\partial \delta
\mathbf{X}}{\partial\emph{\textbf{x}}}\right) .
\end{equation*}%
Moreover, we can write:%
\begin{equation*}
\int_{W}\left[ \rho\, (\overset{}{\overset{\centerdot }{\mathbf{\Xi }%
}^{\tau }-g_{\mathbf{X}}^{\prime })}\delta \mathbf{X}+tr\left( \sigma _{1}F\frac{%
\partial \delta \mathbf{X}}{\partial \emph{\textbf{x}}}\right) \right] \,d%
\emph{\textbf{x}}\,dt=0.
\end{equation*}%
Hence%
\begin{equation*}
\int_{W}\left\{ \rho\, \left[ \overset{}{\overset{\centerdot }{\mathbf{\Xi }%
}^{\tau }-g_{\mathbf{X}}^{\prime }}-\frac{1}{\rho
}\,\func{div}\left( \sigma _{1}F\right) \right]\right\} \,\delta
\mathbf{X}\ d\emph{\textbf{x}}\,dt=0.
\end{equation*}%
We thus obtain%
\begin{equation*}
\overset{\centerdot }{\mathbf{\Xi }}^{\tau }=g_{\mathbf{X}}^{\prime }+\frac{1%
}{\rho }\,\func{div}(\sigma _{1}F).
\end{equation*}%
Let us notice that the divergence operator is applied to a tensor
defined on $\mathcal{D}_{0}$ with values in $\mathcal{D}_{t}$; the
partial derivatives are calculated on $\mathcal{D}_{t}.$

\qquad\qquad\qquad\qquad\qquad\qquad\qquad\qquad\qquad\qquad\qquad\qquad%
\qquad\qquad\qquad\qquad\quad $\Box$

\end{document}